\documentclass[12pt]{iopart}

\usepackage{graphicx}
\usepackage{amssymb}
\newcommand{\bea}{\begin{eqnarray}}
\newcommand{\eea}{\end{eqnarray}}

\begin{document}
\title{Spectral transitions in networks}

\author{Gergely Palla$^{1,3}$ and G\'abor Vattay$^{2,3}$}
\address{$^1$ Biological Physics Research Group of HAS, E\"otv\"os University, P\'azm\'any P.\ stny.\ 1A, H-1117 Budapest, Hungary}
\address{$^2$ Dept. of Complex Systems, E\"otv\"os University, P\'azm\'any P.\ stny.\ 1A, H-1117 Budapest, Hungary}
\address{$^3$ Collegium Budapest Institute for Advanced Study, Szenth\'aroms\'ag u.2, H-1014 Budapest, Hungary}

\begin{abstract}
We study the level spacing distribution $p(s)$ in the spectrum of random
 networks. According to our numerical results, the shape of $p(s)$ in 
 the Erd\H os-R\'enyi (E-R) random graph is  determined by the
 average degree $\left< k\right>$, and $p(s)$ undergoes a dramatic
 change when $\left< k\right>$ is varied around the critical point of
 the percolation transition, $\left< k\right>=1$. 
When $\left< k\right> >> 1$, the $p(s)$
 is described by the statistics of the Gaussian Orthogonal Ensemble (GOE), 
one of the major statistical ensembles in Random Matrix Theory, whereas
 at $\left< k\right>=1$ it follows the Poisson
 level spacing distribution.  Closely
 above the critical point, $p(s)$ can be described in terms of an
 intermediate distribution between Poisson and the GOE, 
the Brody-distribution. Furthermore, below the critical point 
$p(s)$ can be
 given with the help of the regularised Gamma-function.
Motivated by these results, we analyse the behaviour 
 of $p(s)$ in real networks such as the Internet, 
a word association network and a protein protein interaction
 network as well. When the giant component of these networks is destroyed
 in a node deletion process simulating the networks subjected to intentional
 attack, their level spacing distribution undergoes a similar transition
 to that of the E-R graph.
\end{abstract}



\maketitle

\section{Introduction}

A wide class of complex systems occurring from the level of cells to
society can be described in terms of {\it networks} capturing the intricate
web of connections among the units they are made of. 
 Whenever many similar objects  in mutual
 interactions are encountered, these objects can be represented
 as nodes and the interactions as links between the nodes, defining a network.
 The world-wide-web, the science citation index, and biochemical
 reaction pathways in living cells are all good examples of complex systems
 widely modeled with networks, and the set of further phenomena where
the network approach can be used is even more diverse.
Graphs corresponding
 to such real networks exhibit unexpected non-trivial properties,
{\it e.g.}, new kinds of degree distributions, anomalous diameter,
spreading phenomena, clustering coefficient, and correlations
\cite{watts-strogatz,barabasi-albert,albert-revmod,dm-book,barrat}.
In most cases, the overall structure of
 networks reflect the
 characteristic properties of the original systems, and
 enable one to sort seemingly
 very different systems into a few major classes of stochastic graphs
 \cite{albert-revmod,dm-book}.
 These developments have greatly
advanced the potential to interpret the fundamental common features of
such diverse systems as social groups, technological, biological and
other networks.

Another general approach to the analysis of complex systems is provided
 by {\it Random Matrix Theory} (RMT), originally proposed by Wigner and Dyson
 in 1967 for the study of the spectrum of nuclei \cite{wigner}. 
Since then, RMT has been
 successfully used in investigations ranging from 
the studies of phase
 transitions in disordered systems \cite{RMT_phase_tr}, 
through the spectral analysis
 of chaotic systems \cite{bohigas} and the stock market \cite{RMT_stock} 
to the studies of brain responses \cite{RMT_brain}.  Recently, 
the network approach to complex systems and the RMT
 were combined in the analysis of the {\it modular structure} of
 biological networks \cite{luo}. Network modules, also called as communities,
 cohesive groups, clusters, {\it etc.}\ correspond to structural sub-units,
 associated with more highly interconnected parts, with no unique definition
\cite{domany-prl,gn-pnas,zhou,newman-fast,al-parisi,huberman,potts,scott-book,pnas-suppl,everitt-book,knudsen-book,newman-europhys,
our_nature,our_prl}.
Such building blocks (functionally related  proteins \cite{ravasz-science,spirin-pnas}, industrial sectors \cite{onnela-taxonomy},
groups of people \cite{scott-book,watts-dodds}, cooperative players
\cite{play1,play2}, {\it etc.}) can play a crucial role in forming the
 structural and functional properties of the involved networks, therefore
 there has been  a quickly growing interest in the last few years in
 developing efficient methods for locating these modules. One of
 the most well known community finding algorithm today is the Girvan-Newman
 algorithm \cite{gn-pnas,newman-fast}, which is based on recursive 
 deletion of links with the highest
 betweenness. This process leads to splitting of the network to smaller
 parts, corresponding to the communities, and the deletion of the 
 links is stopped, when optimal modularity is reached.

 In the analysis of a protein-protein interaction 
network and a metabolic network, Luo {\it et al.} found that the 
{\it fluctuations of the level spacing in the  spectrum} 
obey different statistics
when the networks are split to the communities given by the 
Girvan-Newman algorithm compared to the original 
state \cite{luo}.
(The spectrum of a network is given  by the eigenvalues
 of its adjacency matrix \cite{mehta,crisanti,illes}). 
For both networks, in the original state the 
fluctuations of the level spacing followed the statistics of 
the {\it Gaussian Orthogonal Ensemble} (GOE), one of the major statistical 
ensembles in RMT. However, when the networks were split to communities,
  the fluctuations in the level spacing became {\it Poissonean}, 
which is another 
important statistics in RMT. Based on this effect, Luo {\it et al.} proposed
 that the monitoring of such changes in the spectral properties 
 can help the identifications of network modules. 

Motivated by these very interesting results, here we study the level
 spacing fluctuations in the spectrum of networks in a more general
 frame work.  Our investigations of the Erd\H{ o}s-R\'enyi (E-R) random graph,
 the Internet, a word association graph and a protein-protein interaction
 graph show that similar spectral transitions occur in
 these networks as well. However, our results indicate that such
 transitions in the spectrum are more likely to be connected to
 the {\it appearance of a giant component} than to the ideal partitioning
 of the network, since {\it e.g.}\ in the E-R graph communities 
 are totally absent. The paper is organised as follows: first
 we summarise the most important properties of the level spacing distribution
 in RMT, then describe our results for the spectral transitions in the
 E-R graph. Finally, we show that similar spectral transitions can be
 induced in real networks as well, simply by destroying the giant component,
 without invoking any sophisticated partitioning of the network to communities.

\section{The level spacing distribution}

The main object of study in RMT is the set of eigenvalues 
$\lbrace e_i \rbrace$ of the random matrix representing the system 
under investigation. In case of networks, this matrix corresponds to
 the adjacency matrix, in which the entry $A_{ij}=1$ if the nodes
 $i$ and $j$ are linked, otherwise $A_{ij}=0$. (For simplicity, let us
 neglect the possible directionality and weight of the links).
One of the most important results of RMT
is that complex systems can be sorted into a few universal classes based 
on the behaviour of the fluctuations in the level spacing between 
these eigenvalues. The level spacing $S$ between two adjacent eigenvalues
 is simply $S_i=e_{i+1}-e_i$, however the distribution of this quantity
 cannot be universal, as there are systems in which eigenvalues are more
 dense/sparse on average compared to others. Therefore, instead the
 the unfolded level spacings are studied, which can
 be defined as
\bea
s_i=\frac{e_{i+1}-e_i}{\left< S \right>_i},
\label{eq:unfold}
\eea
where $\left< S\right>_i$ denotes the local average
 of the level spacing in the vicinity of $e_i$. 
The probability distribution of the unfolded level spacings 
(which from now on we shall call simply as the level spacing distribution)
 can be described with the probability density $p(s)$ and the corresponding
 cumulative distribution $P(s)=\int_0^sp(x)dx$. Due to the unfolding 
(\ref{eq:unfold}), the expectation value of the level spacing is
 one: 
\bea
\left< s\right> =\int sp(s)ds=1.
\label{eq:expect_one}
\eea
The level spacing distribution of systems with strongly correlated
 eigenvalues follows the statistics of the GOE, defined as an ensemble of 
random matrices filled with elements drawn from a Gaussian distribution.
In this case $p(s)$ and $P(s)$ are given by the Wigner-Dyson 
distribution \cite{wigner-dyson} as
\numparts
\bea
p_{\rm GOE}(s)& = &\frac{\pi}{2}s\exp\left(-\frac{\pi}{4}s^2\right), \\
P_{\rm GOE}(s)& = &1-\exp\left(-\frac{\pi}{4}s^2\right).
\label{eq:GOE}
\eea
\endnumparts
Another important universality class is formed by the systems with
no correlation between the eigenvalues, following a Poisson level 
spacing distribution:
\numparts
\bea
p_{0}(s)&=&\exp(-s), \\
P_{0}(s)&=&1-\exp(-s).
\label{eq:Poisson}
\eea
\endnumparts
In chaotic systems with weak disorder, intermediate statistics were 
observed as well, described by the Brody-distribution 
\cite{brody,brody2,robnik,robnik2,karremans}:
\numparts
\bea
p_{\rm B}(s)&=& C\alpha s^{\alpha-1}\exp\left(-Cs^{\alpha}\right),
\label{eq:intera}\\
P_{\rm B}(s)&=&1-\exp\left(-Cs^{\alpha}\right).
\label{eq:interb}
\eea
\endnumparts
where $C$ is a normalising constant ensuring the fulfil of 
Eq.(\ref{eq:expect_one}), and the parameter $\alpha$ determines how
 far the distribution falls from the two limiting cases. 
(At $\alpha=1$ we recover
 the Poisson-distribution, whereas $\alpha=2$ corresponds to
 the statistics of the GOE). In the next section we shall analyse the
 level spacing distribution of the E-R graph.

\section{Spectral transition in the E-R graph}
The concept of random graphs was introduced by Erd\H{o}s and R\'enyi
 \cite{e-r} in the 1950s in a simple model starting 
with $N$ nodes, and connecting
 every pair of nodes independently with the same probability $p$. Even
 though real networks differ from this
 simple model in many aspects, the E-R uncorrelated random graph
remains still of great interest,
since such a graph can serve both
as a test bed for checking all sorts of new ideas concerning
complex networks in general, and
as a prototype of random graphs to which all other random graphs can be
compared. 

Perhaps the most conspicuous early
result on the E-R graphs was related to the percolation transition
taking place at $p=1/N$. 
The appearance of a {\em giant component} in a
network, which is also referred to as the {\em percolating component},
results in a dramatic change in the overall topological features
of the graph and has been in the centre of interest for
other networks as well. The relative size of the largest component
 compared to the total number of nodes is determined by the
 average degree $\left< k\right>=pN$, and the critical
 point of the transition is at $\left< k\right>=1$. 

\begin{figure}[h]
\centerline{\includegraphics[width=0.8\textwidth]{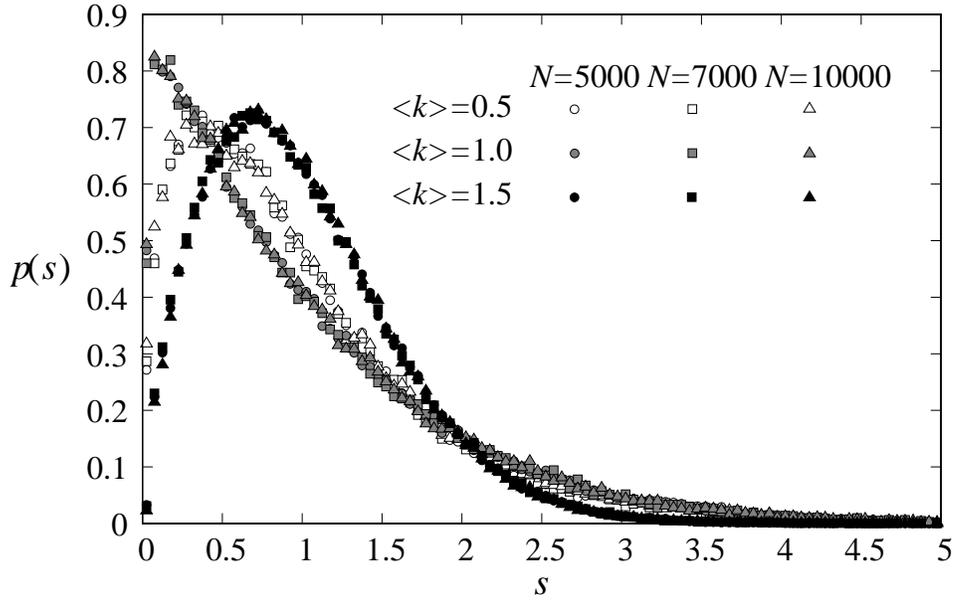}}
\caption{The level spacing distribution $p(s)$ of E-R graphs
of size $N=5000$ (circles), $N=7000$ (squares), and $N=10000$ (triangles) 
 at average degree $\left< k\right>=0.5$ (white symbols), 
$\left< k\right>=1$ (gray symbols), and$\left< k\right>=1.5$ (black symbols).
The curves corresponding to different system sizes with the same average
 coincide with each other.
\label{fig:E-R_below_lsd}}
\end{figure}

In our studies concerning the level spacing distribution of the
E-R graph, we observed a similar phenomenon: the
shape of $p(s)$ is determined by $\left< k\right>$, or in other words,
 the $p(s)$ of E-R graphs with the same average
 degree follow the same curve.
 In Fig.\ref{fig:E-R_below_lsd}. we demonstrate this effect by plotting 
 the level spacing distribution for E-R graphs of size $N=5000$ (circles), 
$N=7000$ (squares) and
 $N=10000$ (triangles), with average degree $\left< k\right>=0.5$ 
(white symbols), $\left< k\right>=1$ (gray symbols),
 and $\left< k\right>=1.5$ (black symbols). 
(For each parameter setting, the spectrum of 
 several different instances of E-R graphs with the given $N$ and 
$\left< k\right>$ was evaluated numerically, and the resulting level 
spacing distributions were averaged). Beside the data collapse for 
the different $N$ parameters, it can be seen that the level spacing
 distribution undergoes a dramatic change when $\left< k \right>$ is 
 varied around $\left< k \right>=1$. The $p(s)$ at 
 $\left< k\right>=1$, the critical point of the 
 percolation transition (denoted by gray symbols) 
 is exponential, whereas it shows a somewhat more complex
 forms for both  $\left< k\right> < 1$ and for $\left< k\right> > 1$. 

\begin{figure}[h]
\centerline{\includegraphics[width=0.8\textwidth]{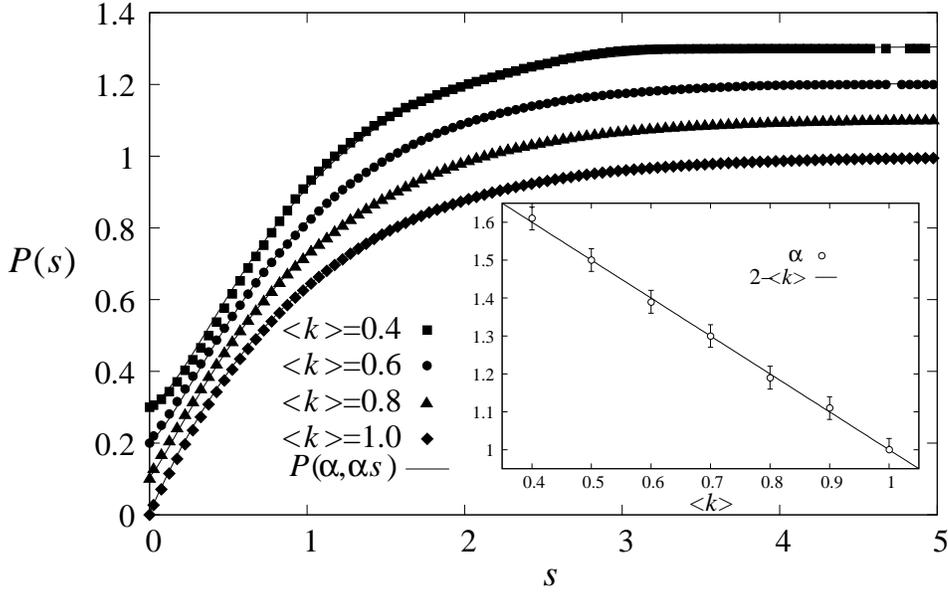}}
\caption{The cumulative level spacing distribution 
$P(s)$ obtained for $\left< k \right>=0.4$ (squares), 
$\left< k\right>=0.6$ (circles), $\left< k\right>=0.8$ (triangles), 
and $\left<k \right>=1$ (diamonds).(The curves corresponding to 
$\left< k\right><1$ were shifted vertically to give a clearer view).
In each case, the empirical $P(s)$ can be very well fitted by 
$P(\alpha,\alpha s)$ (continuous lines). The inset shows the 
fitting parameter $\alpha$ as the function of the average degree 
$\left< k \right>$.
\label{fig:E-R_below_csd}}
\end{figure}

First, let us concentrate on the $\left< k\right> <1$ regime. This corresponds
 to the dispersed state, where the graph consists of small isolated subgraphs.
 In Fig.\ref{fig:E-R_below_csd}. we plot
 the observed cumulative level spacing distribution for $\left< k\right>=0.4,0.6,0.8$ and $\left< k\right>=1$. In each case, the empirical
 results can be very well fitted by 
\numparts
\bea
p(s)&=&\frac{\alpha^\alpha}{\Gamma(\alpha)}\exp(-\alpha s)s^{\alpha-1}\\
P(s)&=&\frac{\gamma(\alpha,\alpha s)}{\Gamma(\alpha)}=P(\alpha,\alpha s),
\label{eq:below_fit}
\eea
\endnumparts
where $\alpha\in [0,1]$ is 
the fitting parameter, $\Gamma(\alpha),\Gamma(\alpha,s)$ and 
$P(\alpha,\alpha s)$ denote the 
Gamma-function, the incomplete Gamma-function and the regularised 
Gamma-function respectively, defined as
\numparts
\bea
\Gamma(\alpha)&=&\int_0^{\infty}t^{\alpha-1}\exp(-t)dt, \\
\gamma(\alpha,s)&=&\int_0^s t^{\alpha-1}\exp(-t)dt, \\
P(\alpha,s)&=&\frac{\gamma(\alpha,s)}{\Gamma(\alpha)}.
\eea
\endnumparts
The distribution given by (\ref{eq:below_fit}) is normalised, and fulfils 
(\ref{eq:expect_one}) as well. The inset in Fig.\ref{fig:E-R_below_csd}. shows the
relation between the fitting parameter $\alpha$ and the average degree,
 which can be expressed simply as
\bea
\alpha=2-\left< k\right>.
\eea
At the critical point of the percolation transition $\alpha$ becomes unity,
  therefore $P(s)$ given by (\ref{eq:below_fit}) is transformed into 
$P_0(s)=-\exp(s)$, corresponding to Poisson statistics in RMT.

\begin{figure}[h]
\centerline{\includegraphics[width=0.9\textwidth]{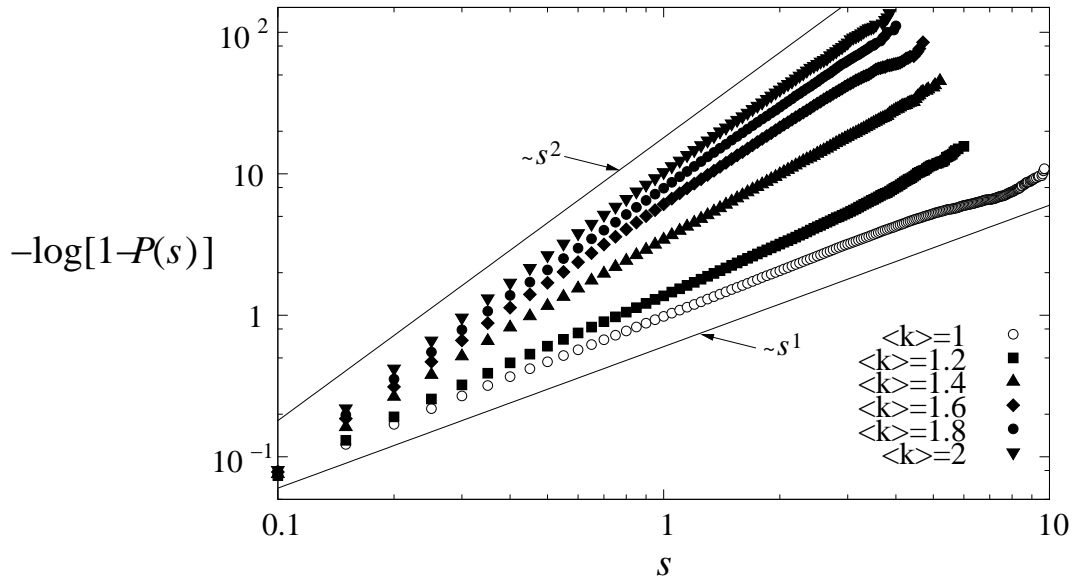}}
\caption{The level spacings of the E-R graph follow the Brody-distribution 
in the $\left< k\right> \geq 1$ regime.
By plotting $-\log[1-P(s)]$ as
 the function of $s$ on logarithmic scale, the $P_{\rm B}(s)$ given
 by Eq.(\ref{eq:interb}) appears as a straight line with slope $\alpha$.
 In the limiting case of 
$\left< k\right>=1$ (open circles) the level spacing
 distribution is Poissonean with $-\log[1-P(s)]=s$, 
corresponding to a straight
 line with unity slope. For $\left<k \right>=1.2$ (squares), 
$\left<k \right>=1.4$ (triangles up), $\left< k\right>=1.6$ (diamonds),
$\left< k\right>=1.8$ (filled circles) and $\left< k\right>=2$ (triangles down)
 we can observe intermediate level spacing distributions between Poissonean
 and GOE, shown by straight lines with increasing slopes.
\label{fig:E-R_inter_brody}}
\end{figure}

\begin{figure}[h]
\centerline{\includegraphics[width=0.8\textwidth]{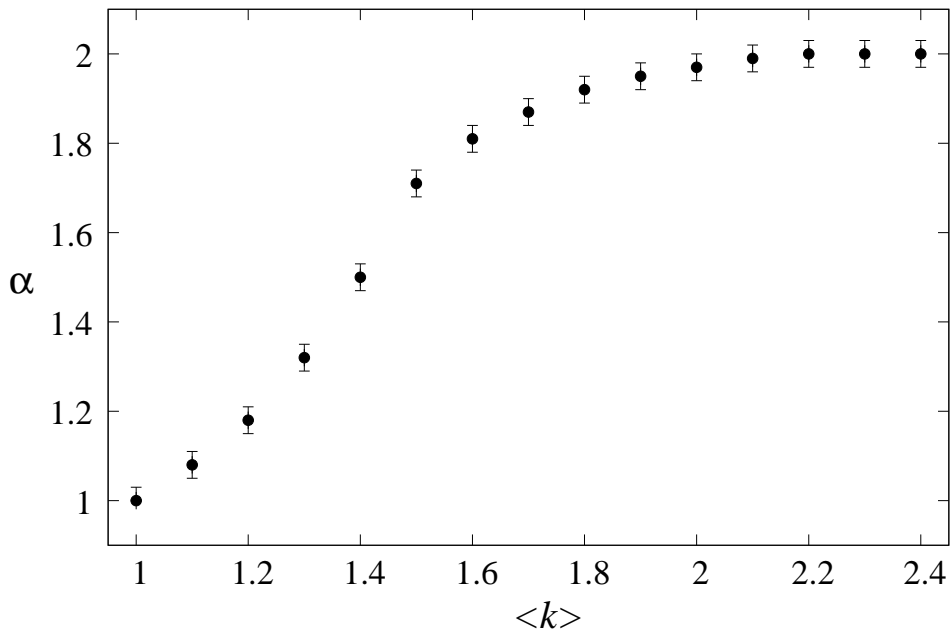}}
\caption{ The fitting parameter $\alpha$ in the $\left< k\right> \geq 1$ regime
 as the function of $\left< k\right>$. Starting from $\alpha=1$ at 
$\left< k \right>=1$, as the level spacing distribution transforms from
 Poissonean to GOE, the parameter $\alpha$ reaches $\alpha=2$ 
following a sigmoid curve.
\label{fig:E-R_inter_fit}}
\end{figure}

In the $\left< k\right> \geq 1$ regime the graph contains a giant component.
Close to the critical point, there are other smaller components present
 as well, however for large enough $\left< k\right>$ the size of 
 the giant component eventually reaches the system size. The level spacing
 distribution in the vicinity of $\left< k\right>=1$ can be fitted with
 the Brody-distribution, given by (\ref{eq:intera}-\ref{eq:interb}), 
 corresponding to a statistics
 in between Poisson and GOE.
 In Fig.\ref{fig:E-R_inter_brody}. we demonstrate this effect by
 plotting $-\log[1-P(s)]$ as the function of $s$ on a logarithmic
 scale. A cumulative level spacing distribution of the form (\ref{eq:interb})
 is thereby transformed into $Cs^{\alpha}$, appearing as a straight line
 with slope $\alpha$. At the critical point of the percolation transition
$\alpha=1$, therefore the slope of the     corresponding curve (open circles)
 is unity. The slope of the curves is increasing with the average degree,
 and at $\left< k\right>=2$ it is already close to $\alpha=2$, corresponding
 to GOE statistics. In Fig.\ref{fig:E-R_inter_fit}. the fitting
 parameter $\alpha$ is shown as the function of $\left< k\right>$, following
 a sigmoid curve, reaching the $\alpha=2$ limit closely above 
$\left< k\right>=2$. 

\section{Spectral transition in real networks}

Similarly to the E-R graph, spectral transitions can occur in real
 networks as well. In our studies we examined the behaviour of the level 
spacing distribution of
 the Internet, a word association network, and a protein-protein interaction
 network. In case of the Internet each node corresponded to an 
Autonomous System, and the links between the Autonomous Systems 
 were obtained from the DIMES project \cite{DIMES}. 
 The word association network was constructed
 from the South Florida Free Association norms list \cite{south-florida}, 
in which a link from 
one word to another indicates that people in the surveys associated the end 
point of the link with its start point. And finally, the studied 
protein-protein network contained the DIP core
list of the protein-protein interactions of \textit{S. cerevisiae}
\cite{dip}. These networks are all scale-free, and they
consist of 14161,
10617, and 2609 nodes and 43430, 63788, and 6355 links, respectively.
In each case, the largest connected component contained more than 90\% of
 the nodes, and the level spacing distribution followed the GOE statistics.

In order to obtain a percolation transition similar
 to that of the E-R graph, we applied the following recursion to all three
 networks until their giant components were destroyed:
\begin{itemize}
\item calculate the node degrees,
\item remove the node with the largest degree.
\end{itemize}
This algorithm is a variation of the method used to investigate
 the attack tolerance of networks, where the nodes are removed in
 the oder of their original degree \cite{attack_tolerance}. Therefore,
 on one hand, the node removal process above can be also viewed as 
the simulation of the intentional damage of the investigated networks.
On the other hand, the advantage of the present approach compared
 to the original process in \cite{attack_tolerance} is that we can 
generate several different
 configurations of the dispersed state:
whenever the largest degree is possessed by more than one node, the
algorithm arrives to a branch point with multiple choices for the next 
node removal. 

\begin{figure}[h]
\centerline{\includegraphics[width=0.8\textwidth]{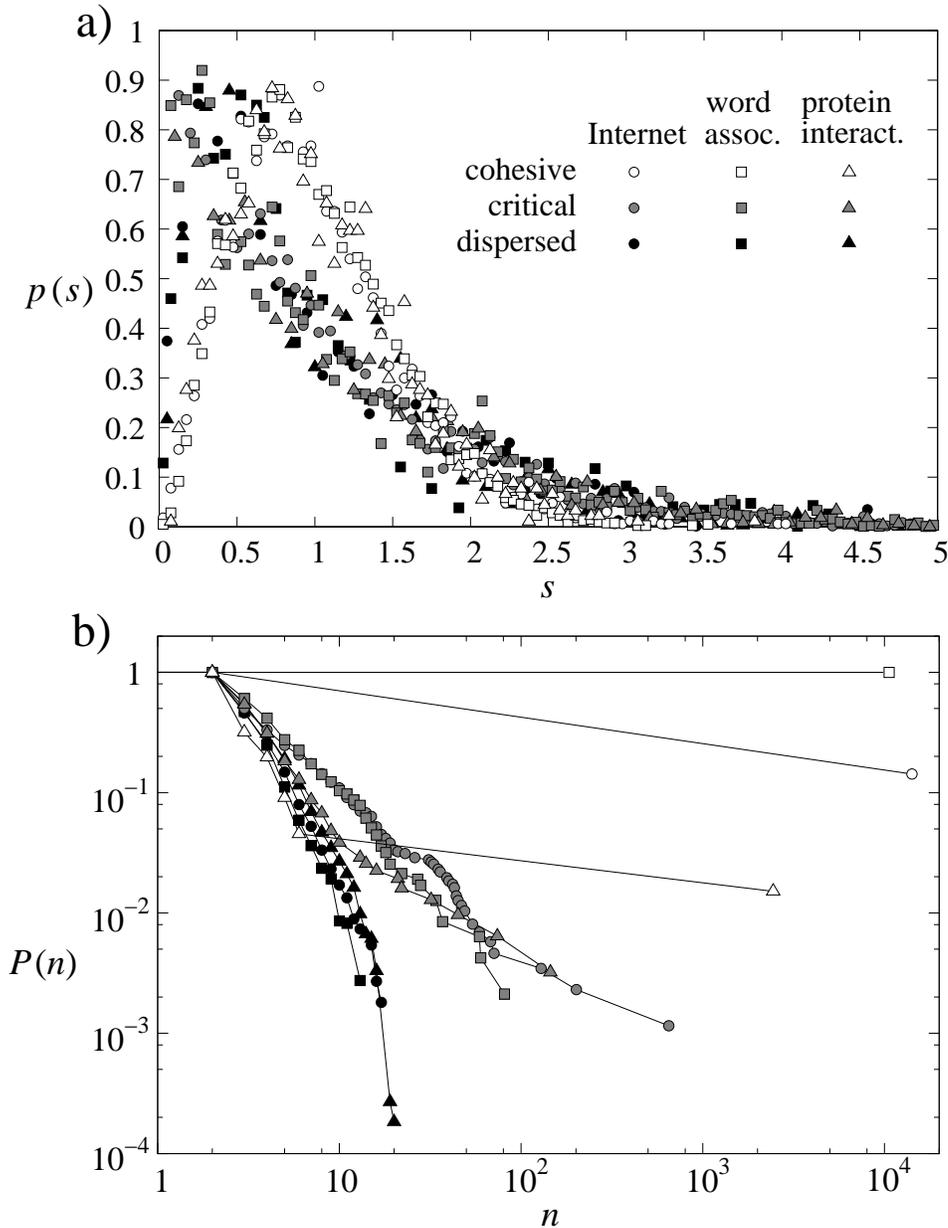}}
\caption{a) The $p(s)$ of the Internet (circles), the word association
 network (squares) and the protein protein interaction network (triangles) 
at three stages in the node deletion process: the original cohesive
 state (white symbols), in the vicinity of the critical point of
 the percolation transition (gray symbols) and in the dispersed state 
(black symbols). 
b) The accompanying cumulative size distributions $P(n)$ with $n$ denoting
 the number of nodes in the components.
\label{fig:real_gr_results}}
\end{figure}


In Fig.\ref{fig:real_gr_results}a we plotted the observed 
level spacing distributions at three stages in the node removal procedure,
 whereas Fig.\ref{fig:real_gr_results}b displays the accompanying cumulative 
component size distributions $P(n)$, where $n$ denotes
 the number of nodes in the components. The circles correspond to
 the Internet, the squares to the word association network, and
 the triangles to the protein-protein interaction network. 
The white symbols show the studied distributions in the original 
cohesive state of the networks: $P(n)$ is dominated 
by an outstandingly large cluster, the giant component, and $p(s)$ follows
 GOE statistics. By succedingly removing 
the nodes with the largest degree, the size of the largest component decreases,
 and in the vicinity of the critical point of the percolation
 transition $P(n)$ is transformed into a power-law, and $p(s)$ becomes 
exponential, as shown by the gray symbols. (These points result from 
averaging over multiple instances of the critical state, 
generated by the algorithm detailed above).
 By continuing the node deletion 
process, the networks fall apart into small disjunct components, $P(n)$ 
transforms into a truncated distribution, and  $p(s)$ becomes peaked again, 
starting from $p(s)=0$ at $s=0$, as shown by the black symbols. (Again,
 the points show the result of averaging over multiple instances of
 the dispersed state).
Even though the $p(s)$ curves for the three different networks do not coincide
 with each other exactly in Fig.\ref{fig:real_gr_results}a, it is clear 
that they all undergo a similar transition to that of the E-R graph.

\section{Conclusions}
According to our investigations the percolation transition in
 networks is accompanied by a transition in the level spacing distribution as 
well. When a giant connected component containing the majority of nodes 
 is present, $p(s)$ follows the GOE
 statistics, whereas in the vicinity of the critical point of the percolation
 transition, $p(s)$ becomes exponential. Dispersed networks
 consisting of many small, disjunct clusters have a $p(s)$ starting from 
$p(s)=0$ at $s=0$ with a peak close to $s=0$, and for the E-R graph, 
the corresponding cumulative level spacing distribution $P(s)$ can 
be simply given by the regularised Gamma-function $P(\alpha,\alpha s)$.

\ack
The authors thank the partial support of the National Science
Foundation (OTKA T37903, F047203), the National Office for Research and
Technology (NKFP 02/032/2004 and NAP 2005/ KCKHA005) and the EU IST
FET Complexity EVERGROW
Integrated Project.

\end{document}